# Energy issues for construction of 10 nm-sized electrostatic trap in saline

*Jingkun GUO, Zijin LEI, Shengyong XU\**

*Key Laboratory for the Physics & Chemistry of Nanodevices, and Department of*

*Electronics, Peking University, Beijing, People's Republic of China*

*\*Author to whom correspondence should be addressed. E-mail: xusy@pku.edu.cn*

**Abstract**

In addition to optical tweezers, magnetic tweezers and dielectrophoresis technique, trapping and manipulating micro/nano-particles with electrostatic tweezers attracted attention in recent years. Here we present a simulation study on the contributions of electrostatic energy, change of entropy, as well as van der Waals interaction in the trapping performance of a 10 nm-sized, pentagon-shaped "nano-trap" for a charged nanoparticle in saline. The results show that the system has a moderate trapping well of moderate depth; van der Waals interaction enhances the trapping capability, and the entropy term induced in distribution contributes significantly in the trapping capability. This work provides some valuable clues to the development of practical sub-micron devices of electrostatic tweezers working in a solution with ions.

Among the well-known four basic physical interactions of the nature, electromagnetic interaction seems the only one that dominates at the micro/nano-scales. Therefore a variety of phenomena related to electromagnetic interactions, such as Coulomb force, Lorenz force, van de Waals force, etc., are among the basic physical terms in micro/nano science and technology, as well as in biosciences at the molecular and sub-cell scales. In several well-developed micro/nano-tweezers, e.g. optical tweezers[1-7], magnetic tweezers[1,8] and dielectrophoresis (DEP) technique[9-11], which are important tools for controlling and manipulating micro/nano-subjects, electromagnetic fields and interactions are directly applied as the working forces. Similarly, in live biosystems, a huge number of effective and regular interactions occur among intracellular micro-/nanoparticles and macromolecules every second[12,13], where the electromagnetic interactions also play the key role[14-17]. However, compared with complicated current artificial current tweezers, with the absence of laser, magnetic field and high frequency field in a natural biosystem, electrostatic force, van der Waals force and entropy factor become the most important mechanisms underlying the vast number of observed complicated facts.

Recently, development of micro/nano-tweezers using electrostatic force has attracted attention. Naturally formed three-dimensional (3D) electrostatic trap was observed in a transmission electron microscopy study[18,19]. The interactions among neutral or charged nanoparticles in an ionic solution became one of the hot topics in soft-material research, results of which might be helpful for construction of trapping and manipulating devices at micro/nanoscales[20,21]. Krishnan et al. constructed a fluidic device and observed effective trapping of nanoparticles with electrostatic charge distributed naturally in nano-sized pits and grooves[20]. At a much larger scale, a prototype device with trap dimension of 30-50 microns was presented to demonstrate basic functions of trapping, releasing and manipulating of charged micro-particles in deionized water[22]. This was a solid step towards the realization of nano-sized devices with full operational functions of electrostatic tweezers which are of great interest for nanoscience and bioscience. But the fabrication of such nano devices remains as a tough technical challenge[23,24].

To check the feasibility for construction of a nano-sized electrostatic trap, in this work we performed simulations on the trapping potential of a 10-nm-sized structure, which has certain pattern of charge distribution but is electrically neutral. Our simulations mainly focused on various energy terms of the system, including electrostatic energy, entropy distribution, total free energy and van der Waals effect. The environment was set as the same saline solution as in a human body, i.e., 0.15 mol/L NaCl in water. To check the validity of the model and formula using in this work, we also calculated the interactions among nano-structures with similar geometric shape but different charge distribution, and the results were consistent with the general phenomena observed in bio-macromolecules.

Figure 1(a) presents schematically the model of the nanostructure used in this work. It is similar to a two-ring structure of charge distribution (see the inset), where the two rings are filled with opposite electrostatic charges, and such a distribution was calculated capable for trapping charged nanoparticles under certain conditions[18]. The continuous two rings are simplified into a pentagram consisting of 10 charged nano-spheres, five positively charged and five negatively charged. With an absolute charge $Q$ of 20$e$, each nano-sphere is set to be 4 nm in diameter. The distance from the center of the pentagon to the center of an outer nano-sphere is defined as $D$ (highlighted in yellow), and all the 5 inner nano-spheres are located at $D/2$ to the center. The whole nanostructure is electrically neutral; therefore it can also be considered as a cluster of 5 dipoles as marked by dashed ovals in Figure 1(a).

Our numerical calculation is based on the nonlinear Poison-Boltzmann equation,

$$\nabla \varphi = \frac{e}{\varepsilon_0 \varepsilon_w} \sum_i z_i c_{i0} \exp\left(-\frac{z_i e \varphi}{k_B T}\right), \qquad (1)$$

where $\varphi$ is potential, $\varepsilon_0$ is the vacuum permittivity, $\varepsilon_w$ is the relative dielectric constant of water (taken as 80 in this work), $k_B T$ is the thermal quantum energy at temperature $T$, and $z_i$ is the charge number of individual ions, i.e., $z_i = 1$ for Na$^+$ and $z_i = -1$ for Cl$^{-1}$ [Ref. 25]. In the saline solution, concentrations of Na$^+$ and Cl$^{-1}$ are much higher than those for H$^+$ and OH$^{-1}$, thus the latter are negligible. The net charge density in solution is thus $\rho = e(c_{Na^+} - c_{Cl^-})$. At locations far away from the nanostructure, the potential of the bulk solution is considered zero; the ion concentration $c_{i0}$ is set as a constant of 0.15 mol/L for both Na$^+$ and Cl$^{-1}$, and $\rho_\infty = 0$. The ion concentration near the nanostructure, $c_i$, is calculated from Boltzmann distribution $c_i(\text{x},\text{y},z) = c_{i0}\exp(-\frac{z_i e \varphi}{k_B T})$. For each nano-sphere, the charges are assumed uniformly distributed with a density of $\rho_{ns}$, leading to $\nabla^2 \varphi = -\frac{\rho_{ns}}{\varepsilon_0 \varepsilon_w}$ inside the nano-sphere.

The equilibrium state of the system is assumed corresponding to the minimum of free energy. The change of Helmholtz free energy $\Delta F$ for the whole system is $\Delta F =$ $U_{el} - T\Delta S$. It consists of an electrostatic energy term $U_{el} = \frac{1}{2}\int \rho\varphi dV$ and an entropy contribution $-T\Delta S$[26,27], where

$$\Delta S = k_B \int \{\sum_i c_0[z_i\psi \exp(-z_i\psi) + \exp(-z_i\psi) - 1]\} dV. \quad (2)$$

Here $\psi = \frac{e\varphi}{k_B T}$ and $k_B$ is the Boltzmann constant.

The values of $U_{el}$, $-T\Delta S$ and $\Delta F$ of the whole system are found very sensitive to $D$. Figure 1(b) presents a typical distribution of $U_{el}$ in the plane of the pentagon nanostructure with $D = 8$ nm. At smaller $D$ values, say, $D = 7.2$ nm, $U_{el}$, $-T\Delta S$ and $\Delta F$ are calculated to be 177.0, 35.8 and 231.3 $k_B T$, respectively. When D increase from 7.2 nm to 10.4 nm, $U_{el}$ in these terms increases but both $\Delta F$ and $-T\Delta S$ decrease, and the differences as compared with that at $D=7.2$ nm are plotted in Figure 1(c). It shows that at $D > 8$ nm, the charging rate of all the three terms get smaller. Therefore, in the following calculation, we fix $D$ at 8 nm for studying the rest properties of the system.

Now we examine the trapping performance of the pentagon nanostructure. Figure 2(a) presents calculated energy depth for the trapping 3 nm diameter nano-sphere with uniformly charged +20 $e$. As defined in Figure 1(a), the pentagon nanostructure is located in the center of $X$-$Y$ plane, and the particle is located at (0, 0, $z$). At $z = 0$, the $U_{el}$, $-T\Delta S$ and $\Delta F$ are calculated to be 203.3, 39.1, and 242.4 $k_B T$, respectively. When $z$ increases, $U_{el}$ decreases, but $-T\Delta S$ increases, and as a result $\Delta F$ still increases. The fitting curve for $\Delta F$ shown in blue indeed clearly shows an effective energy well, for which the trapping depth is around 0.65 $k_B T$ when $z$ is in the region of [-4 nm, + 4 nm]. This characterizes a localized shallow trap.

From the energy terms we can obtain the probability density distribution, by using a general correlation of $f(z) \propto \exp(-\frac{F(z)}{k_B T})$, as plotted in Figure 2(b). In the trapping well, the charged target nanoparticle is supposed to experience a force **T** towards the center, which can be determined by $\mathbf{T}(z) = -\frac{dF(z)}{dz}$. In addition, in the vicinity of $z = 0$, the system can be regarded as a harmonic oscillator[28]. It results in a second-order free energy of $F^{(2)} \approx 0.18 z^2 (k_B T/nm^2)$ and an elastic coefficient of $1.5 \times 10^{-3} N/m$. The data for $F^{(2)}$ is plotted in Figure 2b.

Under this given condition, the absolute value of term $-T\Delta S$ is always found larger than that of $U_{el}$. We attribute it to a strong shielding effect caused by a high numerical density of ions in the system, where the Debye length is in the order of 1 nm. This leads to a weakened effective electrostatic field at locations a few nanometers away from the pentagon nanostructure. Meanwhile, a large number of ions in the solution results in a larger reduction of entropy, because the random distribution of ions in a bulk solution turns into a relatively regular distribution near the pentagon nanostructure.

Another important factor that may affect the performance of the trap is the van der Waals interaction[24]. To calculate the influence of van der Waals effect, the 4-nm diameter nano-spheres and the target nanoparticle are all assumed to be a material with a known dielectric constant. Here two fixed materials are used to obtain the results, silicon and an organic material. The additional van der Waals energy is calculated by using Hamaker equation[29],

$$U(r; R_1, R_2) = -\frac{A}{6}\left(\frac{2R_1 R_2}{r^2-(R_1+R_2)^2} + \frac{2R_1 R_2}{r-(R_1-R_2)^2} + \ln\left[\frac{r^2-(R_1+R_2)^2}{r^2-(R_1-R_2)^2}\right]\right) \quad (3)$$

where $r$ is the distance between two spheres, $R_1$ and $R_2$ are the radius of the two spheres. The Hamaker constant $A$ for silicon and the proposed organic material are set at $0.85\times 10^{-20} J$ and $0.4\times 10^{-20} J$ [24], respectively. The effect of van der Waals interaction is found negligible when *z* is larger than 4 nm. When *z* is approaching zero, the effect becomes stronger, and it enhances the depth of the trap. The effect is found sensitive to the chosen materials of the trap and particle. A system made of silicon has a deeper trap depth than that made of the organic material. The results are plotted in Figure 2(c), which implies that for construction of manmade electrostatic nano-traps, different choices of device materials and different composition of the target nanoparticles may result in remarkable change in the trapping performance.

The trap depth shown in Figure 2(c) ranges from 1-1.5 $k_BT$, depending on the choices of construction material. This may lead to moderate trapping performance, that some fluctuation effects of the surrounding environment, such as the Brownian motion induced by thermal energy, may cause instability of the trapping effect. Another effect to be mentioned is the Stern layer at the solid-liquid interface, which is not included in the present calculation for simplification, may also weaken the performance of the trap[24].

It is interesting to know what may happen if the nano-sphere being trapped by the pentagon nanostructure is replaced by a nano-particle with comparable size to the "trap" itself. We performed simulation on the interaction between two nano-structures with the same geometric shape but different charge distribution by using the same model. To certain level this was done to check the validity of our model. As observed intensively, the interaction strength between two bio-macromolecules could be dramatically influenced by the change of charge distribution in one or both of the two counterparties.

As a mimic to the phase change of bio-macromolecules, we have assumed 5 "dipoles" (highlighted with dot ovals in Figure 1(a)) whose signs can be flipped over, thus it results in 8 different cases as shown in Figure 3(a). Here only a simple configuration is selected for the calculation to give a flavor of the energy terms of the whole two pentagon system. The two pentagons are both parallel to the *X-Y* plane, with one center located at (0, 0, 0) and the other at (0, 0, $z_0$), $z_0$ = 5 nm, and the latter twists an angle in the *X-Y* plane as compared to the former. For such a 5-fold rotational symmetry, the energy terms of $F(\theta)$ are calculated from $\theta = 0°$ to $\theta = 72°$ for all the 8 different configurations (Figure 3 (b)).

Presented in Figure 4(a) and Figure 4(c) are two sets of typical results for the configurations of Case 1 and Case 8 shown in Figure 3(a), respectively. Clearly, for Case 1, the system is more stable at $\theta = 36°$, but for Case 8, the stable location is at $\theta = 0°$. For these two cases, the angle dependent possibility density distribution $f(\theta)$ and torque $\tau(\theta)$ are shown in Figure 4(c) and Figure 4(d), respectively. The van der Waals interaction could also influence the stability of the two-pentagon system. When the contribution of van der Waals interaction is taken into account, the results are presented with dashed green line in Figure 4(a) and Figure 4(c), where constant *A* is taken as $0.4 \times 10^{-20}$ J. In this condition, the van der Waals interaction enhances the angular stability in Case 1, but weakens it in Case 8.

For all different configurations of the two-pentagon combination, it is found that the total system energy is always lower by 10 $k_BT$, or more than the sum of energy when these two pentagons are located far away (infinite) from each other. This indicates a trend of aggregation of these nanostructures. As expected, the results illustrate that with same size and geometric shape, charge distribution of neutral nano-structures can greatly influence the interaction among them.

In short, we have studied the contributions of electrostatic energy, change of entropy and van der Waals interaction in the trapping performance of a 10 nm-sized, pentagon-shaped "nano-trap" for a charged nanoparticle in saline. The results show that under certain configurations the system has a moderate trapping well with a depth of 1.0-1.5 $k_BT$. The van der Waals interaction between the "trap" and target particle enhances the trapping capability, and the strength is material dependent. At this nano-scale and in a saline solution, the contribution from entropy term induced by distribution of ions near the device surface is found playing a remarkable role in the trapping performance. The results of the work offer valuable clues for construction of practical micro/nano-electrostatic tweezers.

**Acknowledgements**

This work is financially supported by NSF of China (Grants 11374016) and MOST of China (Grant 2012CB932702). We thank Miss Jingjing Xu for valuable discussions.

**Figure Captions**

**Figure 1** The schematic diagram of a pentagon nanostructure. The two-ring configuration can also be considered as consisting of 5 "dipoles", which are highlighted with dash loops. The inset is a two-ring structure for an electrostatic trap presented in Ref. 18. (b) The potential distribution of the system in the *X-Y* plane at $z = 0$. (c) The change in $U_{el}$, $-T\Delta S$ and $\Delta F$ as compared to corresponding values calculated at $D = 7.2$ nm.

**Figure 2** (a) Calculated $U_{el}$, $-T\Delta S$ and $\Delta F$ of a pentagon structure at $D = 8$ nm for a target nanoparticle located at $(0, 0, z)$ with a charge of 20 $e$. (b) The probability density distribution of the system along $z$ direction. Insert: second-order polynomial fit of free energy at the vicinity of $z = 0$. (c) The sum value of $\Delta F$ when additional energy induced by van der Waals interaction is taken into account.

**Figure 3**(a) Eight cases of possible combinations of the 5 "dipoles", where each "dipole" has two choices of orientation. (b) The free energy of two-array system corresponding to the eight cases listed in (a).

**Figure 4** (a) Values of $U_{el}$, $-T\Delta S$ and $\Delta F$ for the configurations of Case 1. (b)Probability density distribution for Case 1. (c)Values of $U_{el}$, $-T\Delta S$ and $\Delta F$ for the configurations of Case 8. (d) Probability density distribution for Case 8. The green dash lines in (a) and (c) correspond to the modified $\Delta F$ values when van der Waals interaction of organic material ($A = 0.4 \times 10^{-20}$ J) is taken into account. Inserts in (b) and (d) show the toques of the systems, respectively.

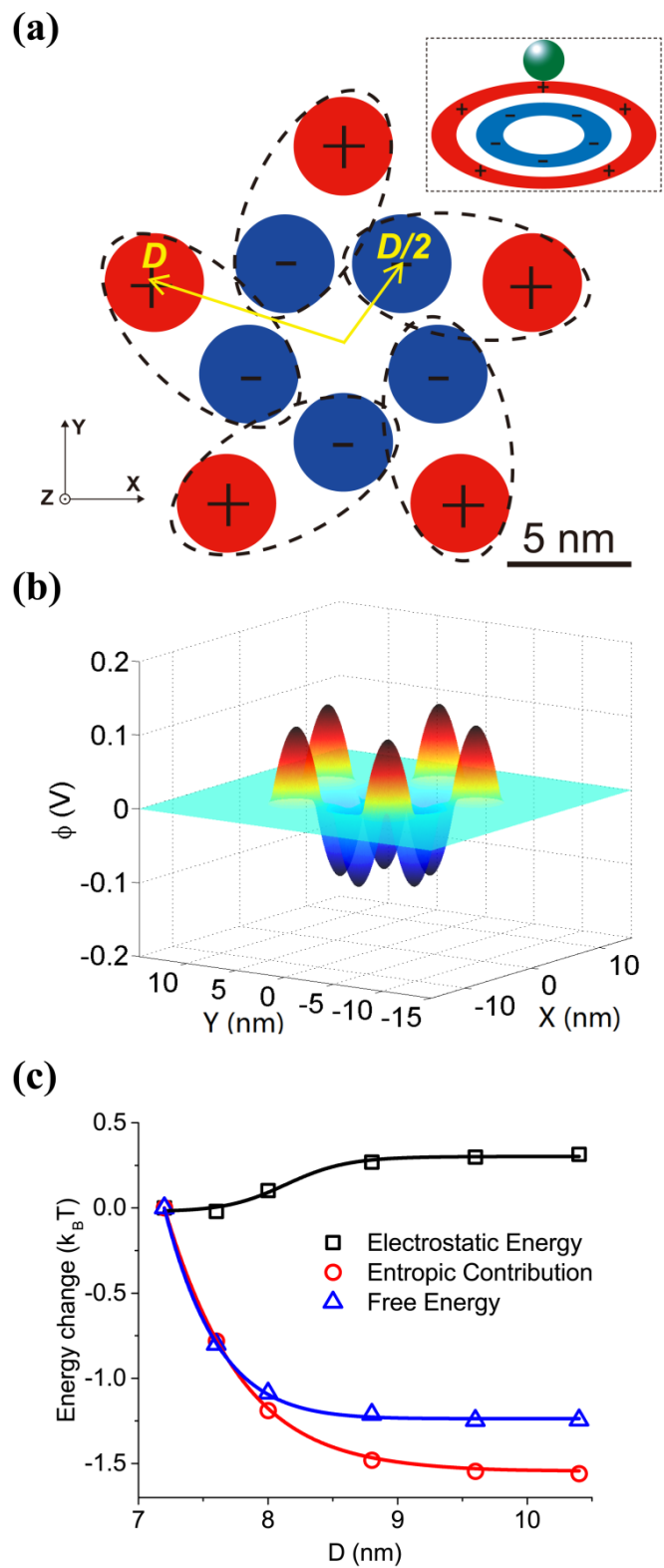

Guo, Lei &Xu, Figure 1

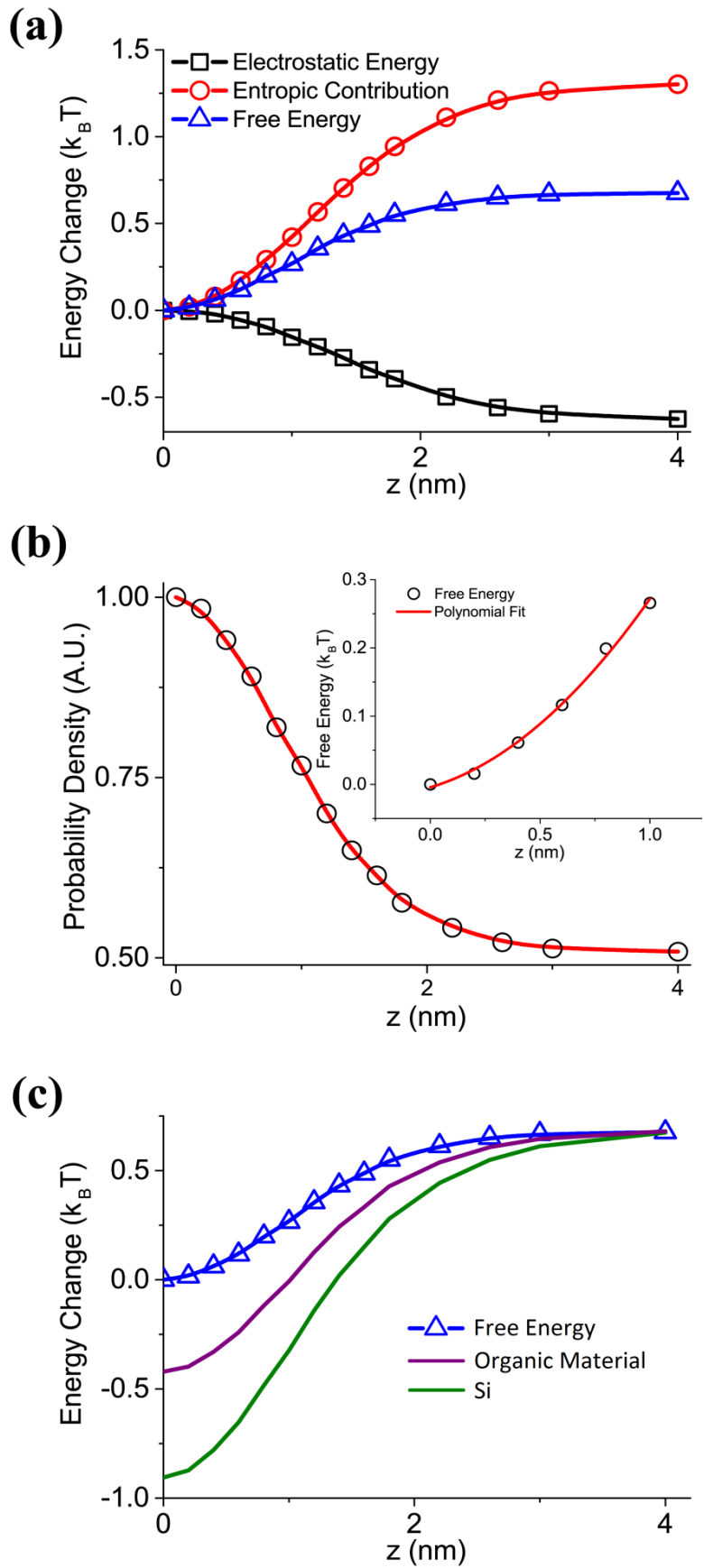

Guo, Lei &Xu, Figure 2

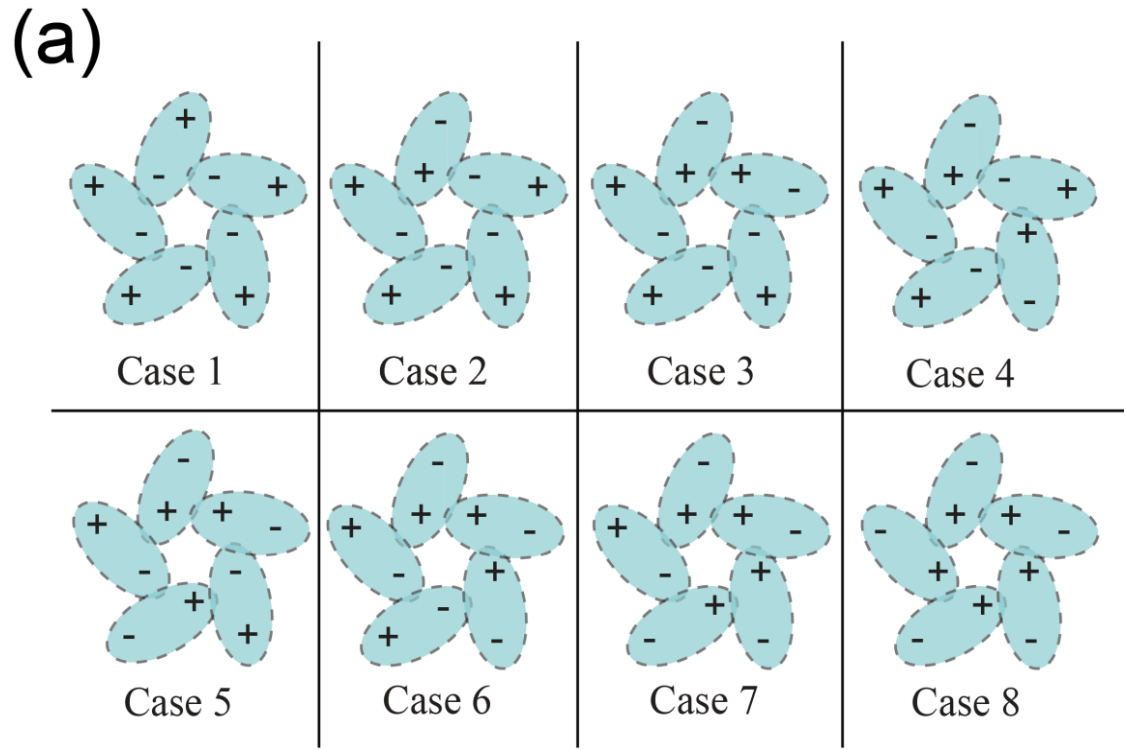

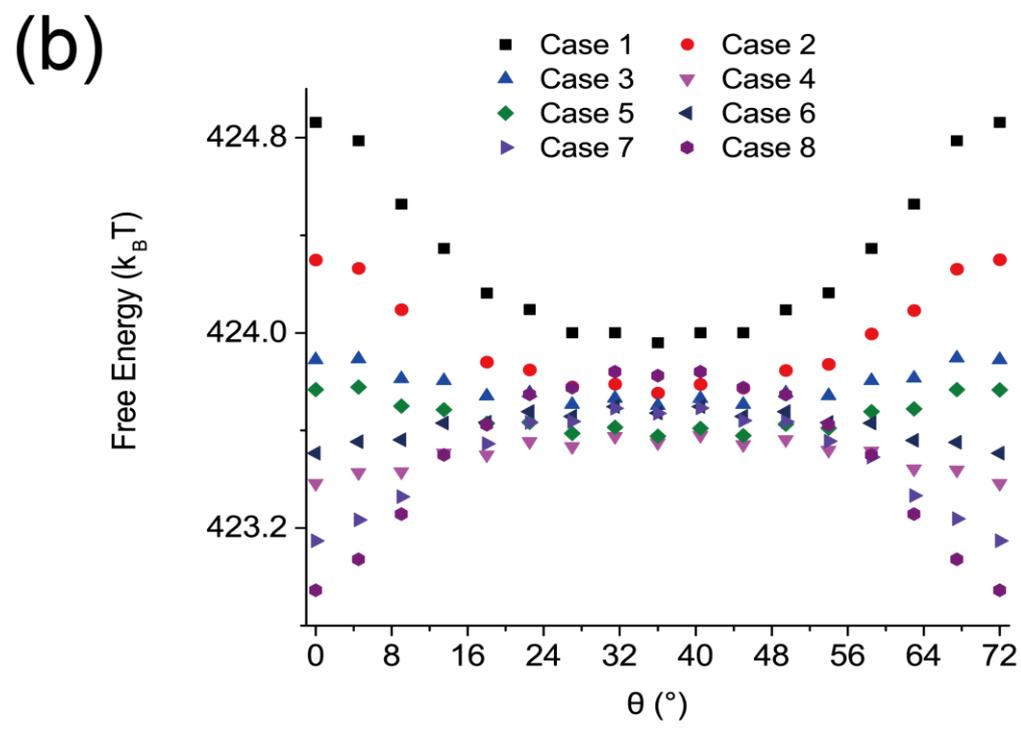

**Guo, Lei & Xu, Figure 3**

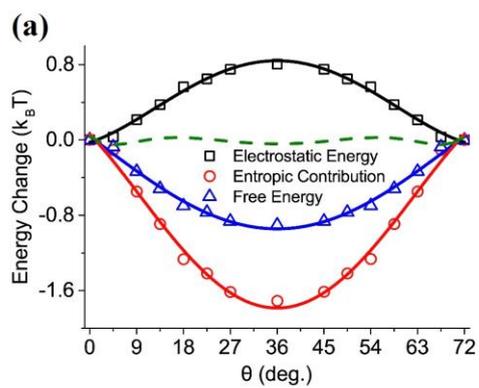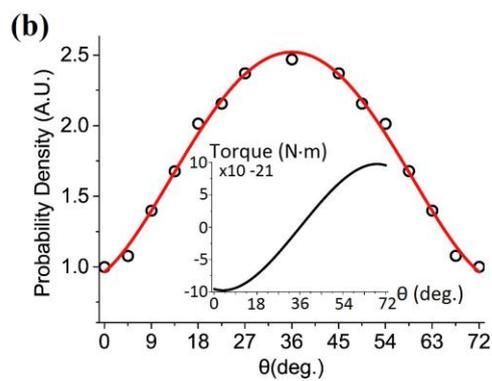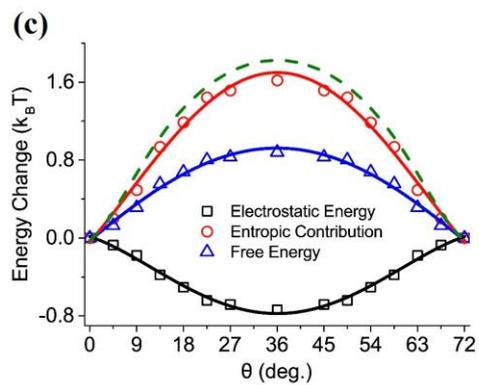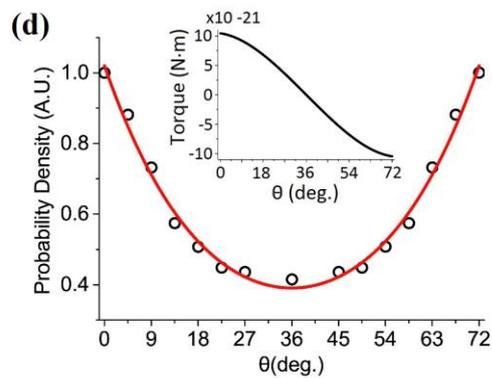

**Guo, Lei & Xu, Figure 4**